\documentclass[12pt]{article}
\usepackage{graphicx}
\begin{document}

\begin{center}
{\bf DIFFRACTIVE DISSOCIATION IN
HIGH ENERGY pp COLLISIONS
IN ADDITIVE QUARK MODEL}

\vspace{.2cm}

Yu.M. Shabelski and A.G. Shuvaev \\

\vspace{.5cm}

Petersburg Nuclear Physics Institute, Kurchatov National
Research Center\\
Gatchina, St. Petersburg 188300, Russia\\
\vskip 0.9 truecm
E-mail: shabelsk@thd.pnpi.spb.ru\\
E-mail: shuvaev@thd.pnpi.spb.ru

\vspace{1.2cm}

\end{center}

\begin{abstract}
\noindent
High energy (CERN SPS and LHC) $pp$ ($p\bar p$) scattering is treated
in the framework of Additive Quark Model together with Pomeron exchange
theory. The reasonable agreement with experimental data is achieved both
for the elastic scattering and for the diffractive dissociation
with natural parameters for the strong matter distribution
inside proton.
\end{abstract}

PACS. 25.75.Dw Particle and resonance production

\section{Introduction}
Regge theory provides a useful tool for phenomenological description
of high energy hadron collisions~\cite{Dr, RMK, MerShab, Sel}.
The quantitative predictions of Regge calculus are essentially dependent
on the assumed coupling of participating hadrons to Pomeron.
In our previous paper~\cite{Shabelski:2014yba} we described
elastic $pp$ ($p\bar p$) scattering including the recent LHC data
in terms of simple Regge exchange approach in the  framework
of Additive Quark Model (AQM)~\cite{LF,VH}.
In the present paper we extend our description to the processes
of single and double diffractive dissociation.

In AQM baryon is treated as a system of
three spatially separated compact objects -- constituent quarks.
Each constituent quark is colored, has internal quark-gluon
structure and finite radius that is much less than the radius of
proton, $r_q^2 \ll r_p^2$. This picture is in good agreement both with
$SU(3)$ symmetry of strong interaction and the quark-gluon structure
of proton~\cite{DDT, Shekhter}.
The constituent quarks play the roles of incident particles
in terms of which $pp$ scattering is described in AQM.

To make the main ingredients and notations of our approach more clear we start from the elastic
scattering in the section~2. The formalism used to describe single and double diffractive
dissociation is presented in the section~3 while the obtained numerical results are compared
with the experimental data in the section~4.

\section{Elastic Scattering Amplitude in AQM}

Elastic amplitudes for the large energy $s=(p_1+p_2)^2$
and small momentum transfer $t$ are dominated by the Pomeron
exchange.
We neglect the small difference in $pp$ and $p\bar p$
scattering coming from the exchange of negative signature Reggeons,
Odderon (see e.g.~\cite{Avila} and references therein),
$\omega$-Reggeon etc, since their contribution is suppressed by $s$.

\begin{figure}[htb]
\centering
\includegraphics[width=0.6\hsize]{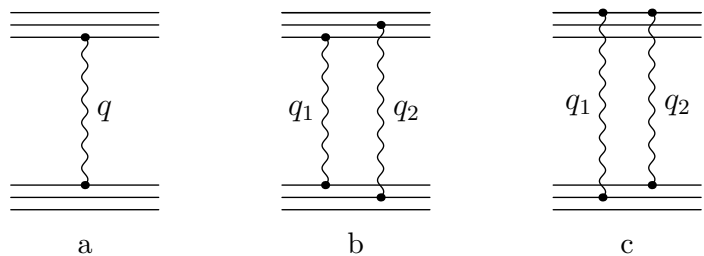}
\caption{\footnotesize The AQM diagrams for $pp$ elastic scattering. The
straight lines stand for quarks, the waved lines denote Pomerons, $Q$ is the
momentum transferred, $t=-Q^2$. Diagram~(a) is the one of the single Pomeron
diagrams, diagrams~(b) and~(c) represent double Pomeron exchange with two
Pomeron coupled to the different quark~(b) and to the same quarks~(c),
$q_1+q_2=Q$.}
\label{1P2P}
\end{figure}

The single $t$-channel exchange results into
amplitude of constituent quarks scattering
\begin{equation}
\label{Mqq} M_{qq}^{(1)}(s,t) = \gamma_{qq}(t) \cdot
\left(\frac{s}{s_0}\right)^{\alpha_P(t) - 1} \cdot
\eta_P(t) \;,
\end{equation}
where $\alpha_P(t) = \alpha_P(0) + \alpha^\prime_P\cdot t$
is the Pomeron trajectory specified by the intercept,
$\alpha_P(0)$, and slope, $\alpha^\prime_P$, values.
The Pomeron signature factor,
$$
\eta_P(t) \,=\, i \,-\, \tan^{-1}
\left(\frac{\pi \alpha_P(t)}2\right),
$$
determines the complex structure of the amplitude. The factor
$\gamma_{qq}(t)=g_1(t)\cdot g_2(t)$ has the meaning
of the Pomeron coupling to the beam and target particles,
the functions $g_{1,2}(t)$ being the vertices of constituent
quark-Pomeron interaction (filled circles in Fig.~\ref{1P2P}).

Due to factorization of longitudinal and transverse degrees
of freedom the longitudinal momenta are integrated over separately
in high energy limit. After this the transverse part of the quark
distribution is actually relevant only.
It is described by the wavefunction
$\psi(k_1,k_2,k_3)$, where $k_i$ are the quark transverse momenta,
normalized as
\begin{equation}
\label{norm}
\int dK\, \bigl|\psi(k_1,k_2,k_3)\bigr|^2\,=\,1,
\end{equation}
and a shorthand notation is used
$$
dK\,\equiv\,d^{\,2}k_1 d^{\,2}k_2 d^{\,2}k_3\,\delta^{(2)}(k_1+k_2+k_3).
$$

The elastic $pp$ (or $p\bar p$, here we do not distinguish between
them) scattering amplitude is basically expressed
in terms of the wavefunction as
\begin{equation}
\label{VQQ}
M_{pp}(s,t)\,=\,\int dK\,dK^\prime
\psi^*(k_i^\prime+Q_i^{\,\prime})\,\psi^*(k_i+Q_i)\,
V(Q,Q^{\,\prime})\,\psi(k_i^\prime)\,\psi(k_i).
\end{equation}
In this formula $\psi(k_i)\equiv \psi(k_1,k_2,k_3)$,
is the initial proton wavefunction
$\psi(k_i+Q_i)\equiv \psi(k_1+Q_1,k_2+Q_2,k_3+Q_3)$
is the wavefunction of the scattered proton,
and the interaction vertex $V(Q,Q^{\,\prime})\equiv
V(Q_1,Q_2,Q_3,Q_1^{\,\prime},Q_2^{\,\prime},Q_3^{\,\prime})$
stands for the multipomeron exchange, $Q_k$ and $Q_l^{\,\prime}$
are the momenta transferred to the target quark $k$ or beam
quark $l$ by the Pomerons attached to them,
$Q$ is the total momentum transferred
in the scattering, $Q^2=-t$.

The scattering amplitude is presented in AQM as a sum over
the terms with a given number of Pomerons,
\begin{equation}
\label{totamp}
M_{pp}(s,t)\,=\,\sum_n M_{pp}^{(n)}(s,t),
\end{equation}
where the amplitudes $M_{pp}^{(n)}$ collect all
diagrams comprising various connections of the beam and target
quark lines with $n$ Pomerons.

Similar to Glauber theory \cite{Glaub, FG} one has to rule out
the multiple interactions between the same quark pair.
AQM permits the Pomeron to connect any two quark lines only once.
It crucially decreases the combinatorics leaving the diagrams
with no more than $n=9$ effective Pomerons.
Several AQM diagrams are shown in Fig.~\ref{1P2P}.

In what follows we assume the Pomeron trajectory in the simplest form
$$
\left(\frac{s}{s_0}\right)^{\alpha_P(t) - 1}\,=\,e^{\Delta\cdot\xi}
e^{-r_q^2\,q^2}, ~~ \xi\equiv \ln\frac{s}{s_0},~~
r_q^2\equiv \alpha^\prime\cdot\xi.
$$
The value $r_q^2$ defines the radius of quark-quark interaction
while $S_0=(9~{\rm GeV})^2$ has the meaning of typical energy scale
in Regge theory.

In the first order there are 9 equal quark-quark
contributions due to one Pomeron exchange between $qq$ pairs.
The amplitude (\ref{VQQ}) reduces to a single term
with $Q_1=Q_1^{\,\prime} = Q$, $Q_{2,3}=Q_{2,3}^{\,\prime} = 0$,
\begin{equation}
\label{M1}
M_{pp}^{(1)}\,=\,9\biggl(\gamma_{qq}\eta_P(t) e^{\Delta\cdot\xi}
\biggr)\,e^{-r_q^2\,Q^2}F_P(Q,0,0)^2,
\end{equation}
expressed through the overlap function
\begin{equation}
\label{FP}
F_P(Q_1,Q_2,Q_3)\,=\,\int dK\,\psi^*(k_1,k_2,k_3)\,
\psi(k_1+Q_1,k_2+Q_2,k_3+Q_3).
\end{equation}
The function $F_P(Q,0,0)$ plays a role of proton formfactor
for the strong interaction in AQM.

An example of the second order diagrams is shown in Fig.~\ref{1P2P}b,c.
Denoting $q_{1,2}$ the transverse momenta carried by
the Pomerons, we have for the diagram~b $Q_1=Q_3^{\,\prime} =0$,
$Q_2=Q_2^{\,\prime} = q_2$, $Q_3=Q_1^{\,\prime} = q_1$
and for the diagram~c $Q_1=q_1+q_2$, $Q_2=Q_3=0$,
$Q_2^{\,\prime} = q_1$, $Q_3^{\,\prime} = 0$.

Generally, the higher orders elastic terms are expressed through
the functions (\ref{FP}) integrated over Pomerons' momenta,
\begin{eqnarray}
\label{Mn} M^{(n)}(s,t)\,&=&\,i^{n-1}\biggl(\gamma_{qq}\eta_P(t_n)
e^{\Delta\cdot\xi}\biggr)^n\,
\int\frac{d^{\,2}q_1}{\pi}\cdots \frac{d^{\,2}q_n}{\pi}
\,\pi\,\delta^{(2)}(q_1+\ldots +\,q_n-Q)\,\\
&&\times\,e^{-r_q^2(q_1^2+\ldots + q_n^2)}\,
\frac 1{n!}\sum\limits_{n~\rm connections}
F_P(Q_1,Q_2,Q_3)\,F_P(Q_1^{\,\prime},Q_2^{\,\prime},Q_3^{\,\prime}),
~~~t_n\simeq t/n. \nonumber
\end{eqnarray}
The sum in this formula refers to all distinct ways
to connect the beam and target quark lines with $n$ Pomerons
in the scattering diagram. The set of momenta $Q_i$ and $Q_l^\prime$
the quarks acquire from the attached Pomerons is particular
for each connection pattern. More detailed description can be found
in \cite{Shabelski:2014yba}.

With the amplitude (\ref{totamp}) the differential cross section
in the normalization adopted here is evaluated as
\begin{equation}
\label{ds/dt} \frac{d\sigma}{dt}\,=\,4\pi\,\bigl|M_{pp}(s,t)
\bigr|^2.
\end{equation}
The optical theorem, that relates the total elastic cross section
and imaginary part of the amplitude, in this normalization reads
$$
\sigma_{pp}^{tot}\,=\,8\pi\,{\rm Im}\, M_{pp}(s,t=0).
$$

Recall once more that exchanges of the positive
signature Reggeons determine, strictly speaking, half of the sum
of $pp$ and $p\bar p$ elastic amplitude. Their difference is neglected
in the present approach.

\section{Cross section of single and double diffractive
dissociation}

\begin{figure}[htb]
\centering
\includegraphics[width=0.6\hsize]{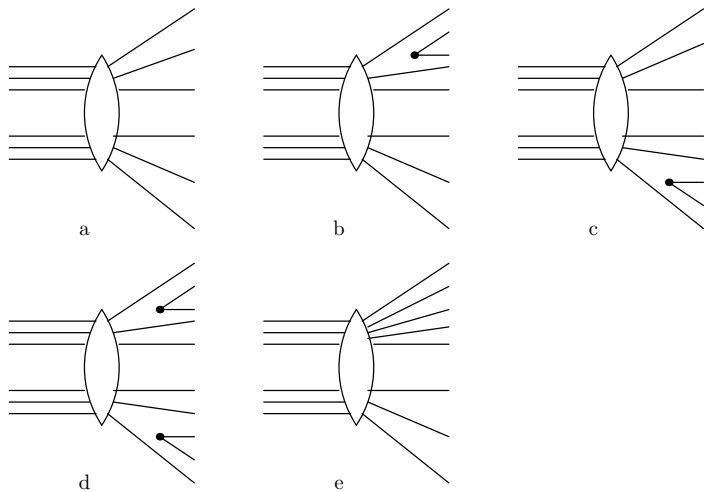}
\caption{\footnotesize Different final states in the high energy
$pp$ collision: a)~elastic $pp$ scattering, b) and c)~single diffractive
dissociation of first or second proton, d)~double diffractive
dissociation, e)~process with one $q\bar q$ pair inelastic production
that does not contribute to the calculated $\sigma_{SD}$ but
can contribute to the experimental value $\sigma_{SD}$.}
\label{FinStates}
\end{figure}

The Glauber theory makes it possible to find as well the cross sections
of excitation or disintegration of one or both colliding objects.
The close approximation (completness condition) (\cite{FG, BraunShabel}
allows one to calculate the total cross sections of all processes
related to the elastic scattering of constituents but without
giving rise to new particles production,
$$
\frac{d\sigma}{dt}(pp\to p^\prime p^\prime) = \frac{d\sigma}{dt}(pp\to pp)
+ 2\frac{d\sigma}{dt}(pp\to p^*p) + \frac{d\sigma}{dt}(pp\to p^*p^*)
$$
Here $d\sigma(pp\to pp)/dt=d\sigma_{el}/dt$ is the
elastic $pp$ scattering cross section shown in Fig.~\ref{FinStates}a.
The situations when one scattered constituent receives
comparatively large transverse momentum is shown in Fig.~\ref{FinStates}b
and Fig.~\ref{FinStates}c. In the case of nucleus-nucleus collision
it results in the excitation or disintegration of one of the nucleus.
In the case of $pp$ collisions in AQM a scattered quark moves
far away from the remnant at the distance $\sim 1$~fm where
a new $q \bar q$ pair is produced due to quark confinement effects.
It can be interpreted as a diffractive production of a single jets,
$\sigma(pp\to pp^*)=\sigma_{SD}$,
say, one or several pions, $p\to p + n\times\pi$.
Similarly two diffractive jet are produced in the case
of Fig~\ref{FinStates}d, $\sigma(pp\to p^*p^*)=\sigma_{DD}$.
In Fig~\ref{FinStates}e a new $q\bar q$ pair is produced
as a part of multipheral ladder independently of
the quarks wavefunction leaving it essentially intact.
This process is related to the inelastic interaction
of the constituents and does not contribute to
$d\sigma(p^{\,\prime} p^{\,\prime})/dt$ in AQM.

The amplitude of single diffraction dissociation reads
\begin{equation}
\label{ASD}
M_{SD}(s,t)\,=\,\int dK\,dK^\prime
\psi^*(k_i^\prime+Q^{\,\prime})\,\widetilde\psi_m^*(k_i+Q_i)
\,V(Q,Q^{\,\prime})\,\psi(k_i^\prime)\,\psi(k_i).
\end{equation}
Here the wavefunction of one of the protons remains unchanged
whereas the other proton turns into $p^*$ final state specified with
the wavefunction $\widetilde\psi_m(k_i)$.

The double diffraction dissociation implies both protons
to be in the $p^*$ final states,
\begin{equation}
\label{ADD}
M_{DD}(s,t)\,=\,\int dK\,dK^\prime
\widetilde\psi_m^*(k_i^\prime+Q^{\,\prime})\,
\widetilde\psi_n^*(k_i+Q_i)\,V(Q,Q^{\,\prime})\,
\psi(k_i^\prime)\,\psi(k_i).
\end{equation}

To obtain cross section one has to square the module of an appropriate
amplitude. Making no distinction between the individual final states
it should be summed up over $m$ for (\ref{ASD}) process
or over $m$ and $n$ indices for (\ref{ADD}) process.
For DD case it gives
\begin{eqnarray}
&&\frac{d\sigma_{el}}{dt}\,+\,2\,\frac{d\sigma_{SD}}{dt}
\,+\,\frac{d\sigma_{DD}}{dt}\\ \nonumber
&=&4\pi\,\sum\limits_{m,n}
\int\!\!dK\,dK^\prime\,dP\,dP^{\,\prime}\,
\widetilde\psi_m^*(k_i^\prime+Q_i^{\,\prime})\,
\widetilde\psi_m^*(k_i+Q_i)\,V(Q,Q^{\,\prime})
\psi(k_i^\prime)\,
\psi(k_i)\\ \nonumber
&&\times\,\psi^*(p_i)\psi^*(p_i^\prime)\,
\,V^*(Q^{\,\prime\prime},Q^{\,\prime\prime\prime})
\widetilde\psi_m(p_i^\prime+Q_i^{\,\prime\prime\prime})
\widetilde\psi_n(p_i+Q_i^{\,\prime\prime}) \nonumber
\end{eqnarray}
Using now the completeness condition,
$$
\sum_n\widetilde\psi_n(p_i+Q_i^{\,\prime\prime})\,
\widetilde\psi_n^*(k_i+Q_i)
\,=\,
\delta^{(2)}(p_i+Q_i^{\,\prime\prime}-k_i-Q_i)
$$
along with the same condition for the index $m$ we get
\begin{eqnarray}
&&\frac{d\sigma_{el}}{dt}\,+\,2\,\frac{d\sigma_{SD}}{dt}
\,+\,\frac{d\sigma_{DD}}{dt}\\
&&=\,\int\!\!dK\,dK^\prime
\psi^*(k_i+Q_i-Q_i^{\,\prime\prime\prime})
\psi^*(k_i^\prime+Q_i^{\,\prime}-Q_i^{\,\prime\prime})
\,V(Q,Q^{\,\prime})
V^*(Q^{\,\prime\prime},Q^{\,\prime\prime\prime})
\psi(k_i^\prime)\psi(k_i).\nonumber
\end{eqnarray}
The double diffractive dissociation cross section
is provided by the two sets of the Pomeron
exchange diagrams separately associated with
the "left" interaction vertex $V(Q,Q^{\,\prime})$ and
the "right" one $V^*(Q^{\,\prime\prime},Q^{\,\prime\prime\prime})$.
They are summed up independently over the total numbers
of the Pomerons participating in the diagram "from the left"
or "from the right", $m, n =1,9$,
$$
\frac{d\sigma_{el}}{dt}\,+\,2\,\frac{d\sigma_{SD}}{dt}
\,+\,\frac{d\sigma_{DD}}{dt}\,=\,
\sum_{m,n} |M_{p^{\,\prime} p^{\,\prime}}^{(m,n)}|^2(s,t).
$$
Substituting here the Pomeron-quark vertex and introduced
above overlap function (\ref{FP}), we express each term
in the sum as
\begin{eqnarray}
\label{MDD}
&&|M_{p^{\,\prime} p^{\,\prime}}^{(m,n)}|^2\,=\,
\biggl(\gamma_{qq}e^{\Delta\cdot\xi}\biggr)^{m+n}
\bigl[\,i\eta_P(t_m)\bigr]^m\,\bigl[-i\eta_P^*(t_n)\bigr]^n\,\\
&&\times\,\int\frac{d^{\,2}q_1}{\pi}\cdots \frac{d^{\,2}q_m}{\pi}
\,\pi\,\delta^{(2)}(q_1+\ldots +\,q_m-Q)\,\nonumber \\
&&\times\,\frac{d^{\,2}q_{m+1}}{\pi}\cdots \frac{d^{\,2}q_{m+n}}{\pi}
\,\pi\,\delta^{(2)}(q_{m+1}+\ldots +\,q_{m+n}-Q)\,\nonumber \\
&&\times\,e^{-r_q^2(q_1^2+\ldots + q_{m+n}^2)}\,
\frac 1{m!}\,\frac 1{n!}\sum\limits_{m,n~\rm connections}
F_P(Q_1,Q_2,Q_3)\,F_P(Q_1^{\,\prime},Q_2^{\,\prime},Q_3^{\,\prime}).
\nonumber
\end{eqnarray}

For the single diffractive dissociation the cross section
includes the single sum over the final states of one of the protons.
In this case the completeness condition gives
$$
\frac{d\sigma_{SD}}{dt}\,+\,\frac{d\sigma_{el}}{dt}\,=\,
4\pi\sum_{m,n} |M_{p p^{\,\prime}}^{(m,n)}|^{\,2}(s,t),
$$
with
\begin{eqnarray}
\label{MSD}
&&|M_{p p^{\,\prime}}^{(m,n)}|^2\,=\,
\biggl(\gamma_{qq}e^{\Delta\cdot\xi}\biggr)^{m+n}
\bigl[\,i\eta_P(t_m)\bigr]^m\,\bigl[-i\eta_P^*(t_n)\bigr]^n\,\\
&&\times\,\int\frac{d^{\,2}q_1}{\pi}\cdots \frac{d^{\,2}q_m}{\pi}
\,\pi\,\delta^{(2)}(q_1+\ldots +\,q_m-Q)\,\nonumber \\
&&\times\,\frac{d^{\,2}q_{m+1}}{\pi}\cdots \frac{d^{\,2}q_{m+n}}{\pi}
\,\pi\,\delta^{(2)}(q_{m+1}+\ldots +\,q_{m+n}-Q)\,
e^{-r_q^2(q_1^2+\ldots + q_{m+n}^2)}\,\nonumber \\
&&\times\,
\frac 1{m!}\frac 1{n!}\sum\limits_{m,n~\rm connections}
F_P(Q_1,Q_2,Q_3)\,F_P(Q_1^{\,\prime},Q_2^{\,\prime},Q_3^{\,\prime})\,
\,F_P(Q_1^{\,\prime\prime},Q_2^{\,\prime\prime},Q_3^{\,\prime\prime})
\nonumber
\end{eqnarray}
The first function $F_P$ appears here from the completeness condition
written for the dissociating proton while two other $F_P$ functions
in the product stand for elastically scattered proton.

\section{Numerical calculations}

The expressions (\ref{MDD},\ref{MSD}) have been used for the numerical
calculation of the single and double diffractive dissociation
of the protons together with the formula (\ref{Mn}) for
their elastic scattering. The overlap function (\ref{FP})
is evaluated through the transverse part of the quarks'
wavefunction, which has been taken in a simple form of two
gaussian packets,
\begin{equation}
\label{gausspack}
\psi(k_1,k_2,k_3)\,=\,N\bigl[\,e^{-a_1(k_1^2+k_2^2+k_3^2)}\,
+\,C\,e^{-a_2(k_1^2+k_2^2+k_3^2)}\bigr],
\end{equation}
normalized to unity (\ref{norm}). One packet parametrization,
$C=0$, is insufficient to reproduce the experimental data
on elastic scattering \cite{Shabelski:2014yba}
as imposing too strong mutual dependence between the total
cross section, the minimum position in $d\sigma_{el}/dt$
and the value of the slope at $t=0$.

All parameters used in the calculation naturally fall into
two different kinds: the parameters of the Pomeron and those
specifying the structure of colliding particles.
The former type, $\Delta$, $\alpha^\prime$, $\gamma_{qq}$,
refers to the high energy scattering theory while the latter,
$a_{1,2}$ and $C$, details the matter distribution inside
the proton in the low energy limit(similar to density distribution
in atomic nuclear).

We recalculate the elastic scattering cross section
$d\sigma_{el}/dt$, obtained in our previous paper
\cite{Shabelski:2014yba} assuming the argument
of the signature factor $t_n=t/n=-Q^2/n$
that is natural for Gaussian $Q^2$ dependence.
It causes a slight change of the model parameters.
Now the Pomeron parameters are
$$
\Delta=0.107,~~~
\alpha^\prime=0.31\,{\rm GeV}^{-2},~~~
\gamma_{qq}=0.44\,{\rm GeV}^{-2},
$$
and the parameters of matter distribution in the proton are
$$
a_1=4.8\,{\rm GeV}^{-2},~~~
a_2=1.02\,{\rm GeV}^{-2},~~~
C=0.133.
$$
Note that the same set of the Pomeron parameters
describes proton and antiproton scattering, therefore
both $pp$ and $p\bar p$ data have been commonly used to fix their values.

\begin{figure}[htb]
\vskip -1.cm
\includegraphics[width=.45\textwidth]{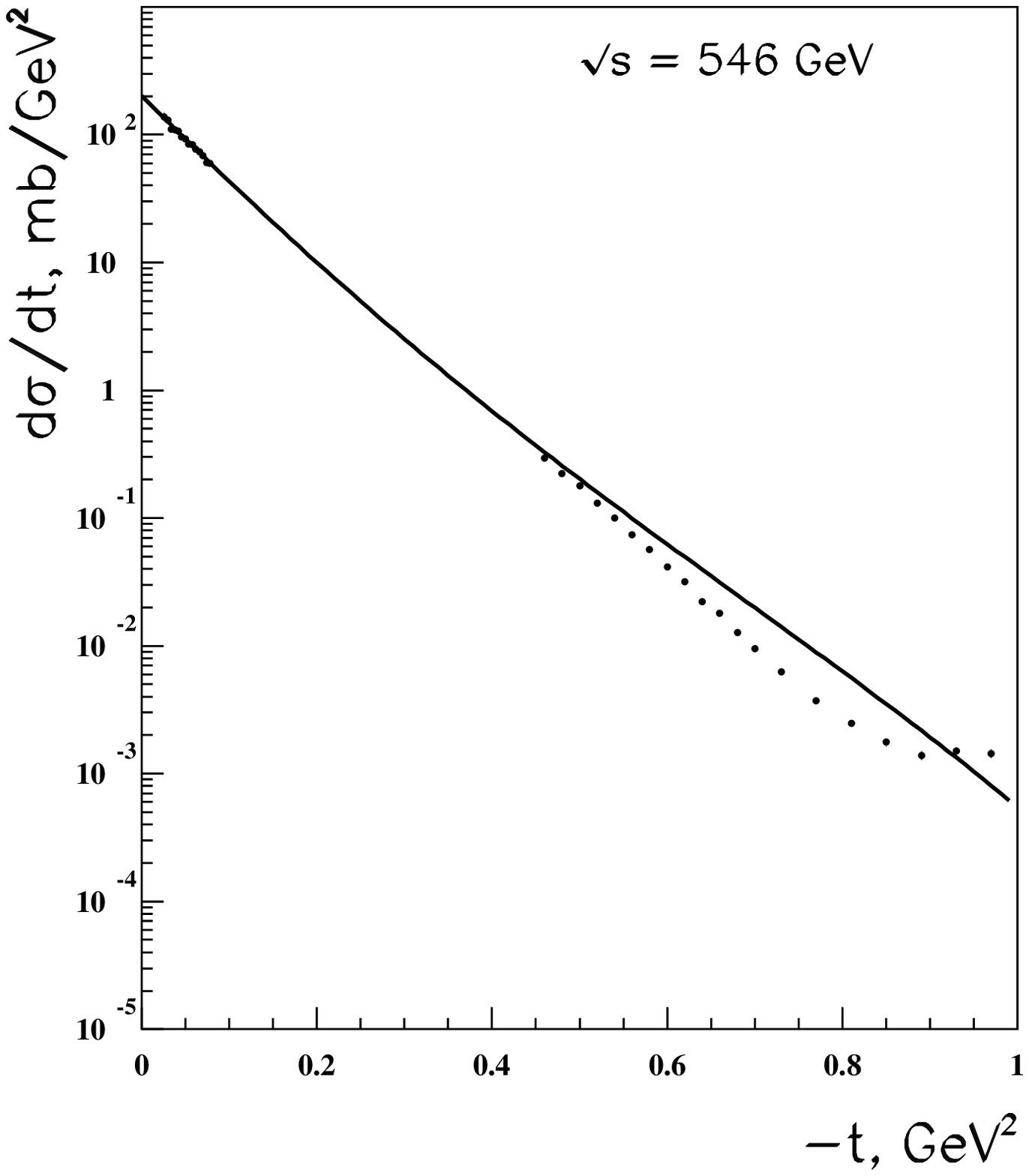}
\includegraphics[width=.45\textwidth]{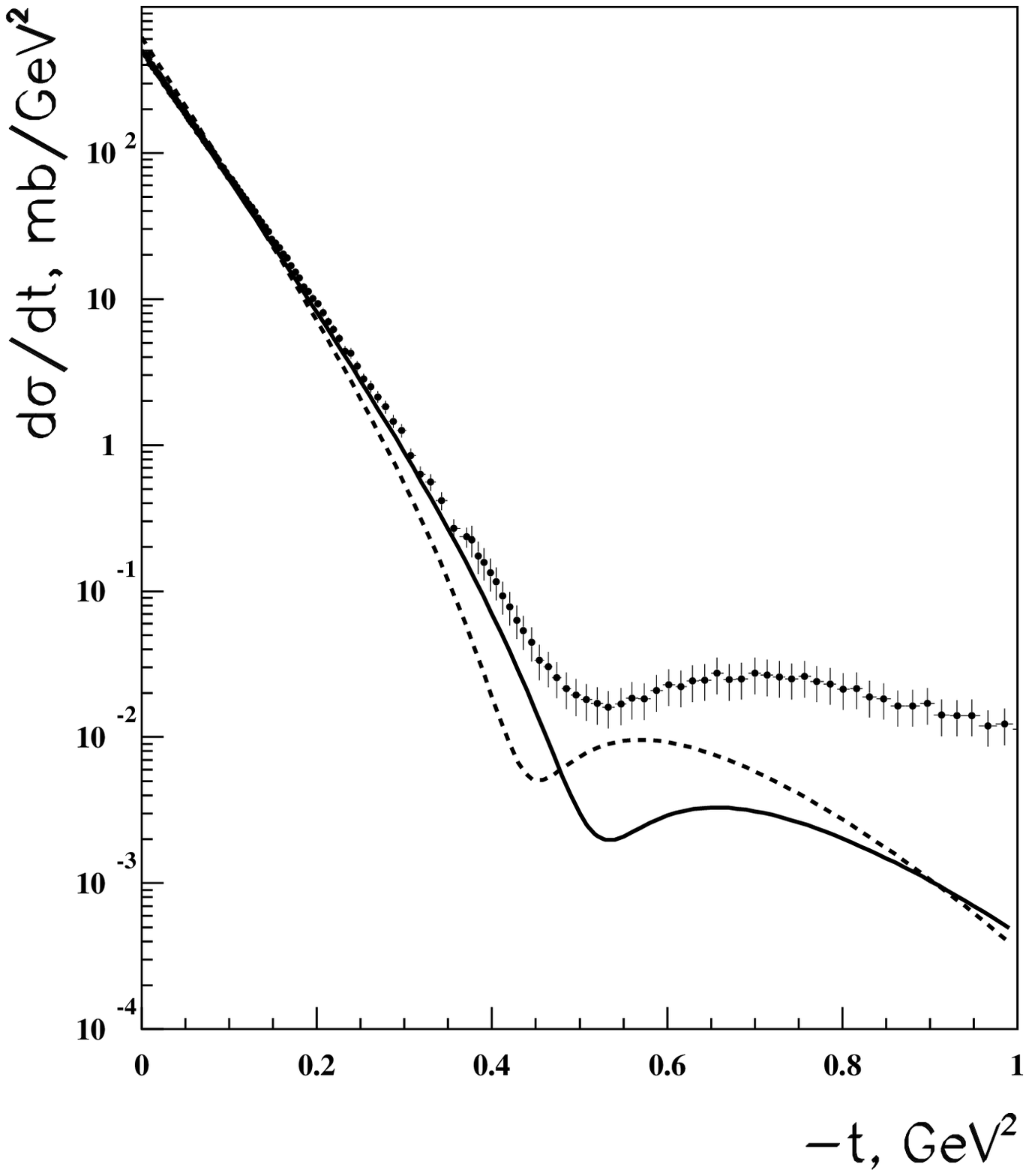}
\vskip -1.5 cm
\caption{\footnotesize
The differential cross section of elastic $p\bar p$ scattering
at $\sqrt s = 546$~GeV (left panel) and for the elastic $pp$ collisions
at $\sqrt s = 7$~TeV (right panel, solid line) compared to the experimental
data. The dotted line at the right panel shows the predicted elastic $pp$
cross section at $\sqrt s = 13$~TeV.
The experimental points have been taken from
\cite{Abe:1993xx,TOT1a,TO1,TO2}.
}
\label{elastic}
\end{figure}

The model gives a reasonable description of elastic scattering
experimental data both for $pp$ collisions at $\sqrt{s}=7$~TeV
and $p\bar p$ collisions at $\sqrt{s}=546$~GeV, see Fig.~\ref{elastic}.
The obtained curves only are slightly different from those published
in the previous paper \cite{Shabelski:2014yba}.

The results for the SD and DD cross sections are presented
in the Table~1. The SD cross sections come out to be rather
small, $\sigma_{SD}/\sigma_{el}\simeq 15 - 18\%$,
that matches perhaps the experimental results at LHC energies
\cite{Antchev:2013haa, OO, Khoze:2014aca}.
The total diffraction cross section is approximately half the elastic one,
$2\sigma_{SD}+\sigma_{DD}\simeq \sigma_{el}/2$,
within the range of available energy
dependence of the probability of diffractive to elastic scattering.

The ratio $\sigma_{SD}/\sigma_{el}$ is in somewhat inconsistency
($1.5 \div 2$ times lesser) with the intermediate energy estimation
in \cite{Kaidalov}. We get $\sigma_{SD}\approx \sigma_{DD}$, so that
$\sigma_{DD}/\sigma_{el}$ is not quadratically small compared
to $\sigma_{SD}/\sigma_{el}$.
The reason for this comes in AQM from an extra third formfactor
$F_P$ in the SD cross section (\ref{MSD}) compared
to the two formfactors in the DD formula (\ref{MDD}).
On the other hand the connection between diffractive
cross section calculated in AQM and the experimental data
is not straightforward since AQM comprises only a part
of the processes involved in the scattering.
The processes shown in Fig.~\ref{FinStates}e are not accounted for
in AQM although their contribution to the experimentally measured
$\sigma_{SD}$ is quite possible.

\noindent
Table 1.
\begin{center}
\begin{tabular}{|c||c|c|c|} \hline
$\sqrt{s}$  & $\sigma_{el}$ (mb) & $\sigma_{SD} (mb) $ &
$\sigma_{DD} (mb) $\\
\hline
546 GeV & 14.3 & 2.3 & 2.6 \\
\hline
7 TeV & 27.3 & 4.3 & 3.9 \\
\hline
13 TeV & 31.6 & 5.4 & 4.9 \\
\hline
\end{tabular}
\end{center}

Motivated by the recently announced new LHC run we present also
the predictions for the elastic $pp$ scattering
and diffractive dissociation
at $\sqrt{s}=13$~TeV. In particular, we expect the total cross
section $\sigma(pp)_{tot}=110$~mb, the parameter of the elastic
slope cone $B=21.8$, the minimum position at $|t|=0.45$~GeV$^2$
while our results for the differential cross
section, $d\sigma_{el}/dt$, are shown in Fig.~\ref{elastic}.

\begin{figure}[htb]
\vskip -1.cm
\includegraphics[width=.45\textwidth]{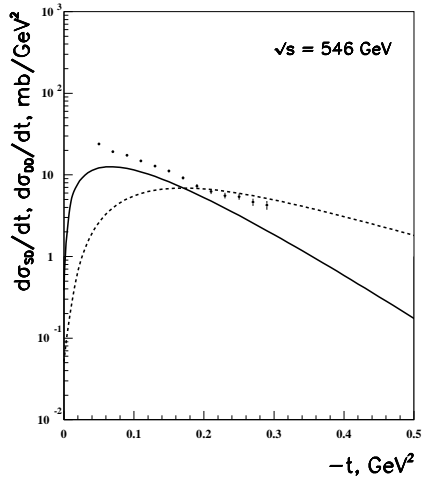}
\vskip -1.5 cm
\caption{\footnotesize The cross section of single (solid line)
and double (dotted line) diffractive dissociation in
$p\bar p$ scattering at $\sqrt s = 546$~GeV.
The experimental SD points have been taken from
\cite{Bernard:1986yh}. }
\label{ppSDDD}
\end{figure}

Fig.~\ref{ppSDDD} shows our results for the differential
cross sections $d\sigma_{SD}/dt$ and $d\sigma_{DD}/dt$ at
$\sqrt s = 546$~GeV.
Unfortunately we are unable to predict
at small $|t|<0.1$~Gev$^2$ because of the unknown
effects of confinement that could lead to the transition between
the ground and excited states. However such a transition
can not change the total cross sections (\ref{MDD}), (\ref{MSD})
calculated with the help of the close approximation.
The region $|t|>1$~Gev$^2$ is beyond
the reach of our model as well since the internal structure
of the constituent quarks can not be more ignored there.
The diffractive cross section behavior in the intermediate interval
is in reasonable agreement with the experimental data.

\section{Conclusion}

We have presented the unified description of the elastic and diffractive
$pp$ ($p\bar p$) scattering in the framework of AQM.
The main feature of our model is a common set of parameters it employs.
After fitting the parameters that describe the elastic scattering
no more ones have been added for the diffractive case.
The parameters chosen are mainly determined by the structure of the quark
wavefunction for the initial state. Despite this the model yields reasonable
calculated values both for the SD and DD cross sections that can be recognized
as a one further argument in favor of AQM applicability for the high energy
$pp$ scattering.

The authors are grateful to M.G.~Ryskin for helpful discussion.


\begin{thebibliography}{**}

\bibitem{Dr} I.~M.~Dremin,
Phys.\ Usp.\  {\bf 56} (2013) 3 [Usp.\ Fiz.\ Nauk {\bf 183} (2013)
3] [arXiv:1206.5474 [hep-ph]].

\bibitem{RMK} M.~G.~Ryskin, A.~D.~Martin and V.~A.~Khoze,
Eur.\ Phys.\ J.\ C {\bf 72} (2012) 1937

\bibitem{MerShab} C.~Merino and Y.~.M.~Shabelski,
JHEP {\bf 1205} (2012) 013 [arXiv:1204.0769 [hep-ph]].

\bibitem{Sel} O.~V.~Selyugin,
Eur.\ Phys.\ J.\ C {\bf 72} (2012) 2073 [arXiv:1201.4458 [hep-ph]].

\bibitem{Shabelski:2014yba}
Y.~M.~Shabelski and A.~G.~Shuvaev,
JHEP {\bf 1411} (2014) 023
[arXiv:1406.1421 [hep-ph]].

\bibitem{LF} E.~M.~Levin and L.~L.~Frankfurt,
JETP Lett. {\bf 2} (1965) 65.

\bibitem{VH} J.~J.~J.~Kokkedee and L.~Van Hove,
Nuovo Cim.\  {\bf 42} (1966) 711.

\bibitem{DDT}
Y.~L.~Dokshitzer, D.~Diakonov and S.~I.~Troian,
Phys.\ Rept.\  {\bf 58}, 269 (1980).

\bibitem{Shekhter} V.~M.~Shekhter, Yad.Fiz. {\bf 33} (1981) 817;
Sov. J. Nucl. Phys. {\bf 33} (1981) 426.

\bibitem{Avila}
R.~Avila, P.~Gauron and B.~Nicolescu,
Eur.\ Phys.\ J.\ C {\bf 49}, 581 (2007) [hep-ph/0607089].

\bibitem{Glaub} R.~J.~Glauber. In "Lectures in Theoretical Physics",
Eds. W.~E.~Brittin etal., New York (1959), vol.1, p.315.

\bibitem{FG} V.~Franco and R.~J.~Glauber, Phys.Rev. {\bf 142}
(1966) 1195.

\bibitem{BraunShabel}
V.~M.~Braun and Y.~M.~Shabelski,
Sov.\ J.\ Nucl.\ Phys.\  {\bf 35} (1982) 731
[Yad.\ Fiz.\  {\bf 35}(1982) 1247].

\bibitem{Abe:1993xx}
F.~Abe {\it et al.}  [CDF Collaboration],
Phys.\ Rev.\ D {\bf 50} (1994) 5518.

\bibitem{TOT1a} G. Antchev et al., TOTEM Collaboration, Europhys. Lett.
{\bf 96}, 21002 (2011).

\bibitem{TO1} TOTEM Collaboration, G. Antchev et al.,
Europhys.Lett. {\bf 101} (2013) 21002.

\bibitem{TO2} TOTEM Collaboration, G. Antchev et al.,
Europhys.Lett. {\bf 95} (2011) 41001, [arXiv:1110.1385].

\bibitem{Antchev:2013haa}
G.~Antchev {\it et al.}  [TOTEM Collaboration],
Europhys.\ Lett.\  {\bf 101} (2013) 21003.

\bibitem{OO}
TOTEM coll. F. Oljemark and K. Osterberg "Studies of soft single
diffraction with TOTEM at $\sqrt{s}=7$~TeV"
LHC students poster session, 13 March 2013.

\bibitem{Khoze:2014aca}
V.~A.~Khoze, A.~D.~Martin and M.~G.~Ryskin,
Int.\ J.\ Mod.\ Phys.\ A {\bf 30} (2015) 08,  1542004,
[arXiv:1402.2778 [hep-ph]].

\bibitem{Kaidalov}
A.~B.~Kaidalov,
Phys.\ Rept.\  {\bf 50} (1979) 157.

\bibitem{Bernard:1986yh}
D.~Bernard {\it et al.}  [UA4 Collaboration],
Phys.\ Lett.\ B {\bf 186} (1987) 227.

\end{thebibliography}
\end{document}